\documentstyle[12pt]{article}
\def\abstract#1{\vskip 7mm 
	\begin{center}{\large Abstract}\par \bigskip
		\begin{minipage}[c]{12cm}
			\small #1
		\end{minipage}
	\end{center}
}
\def\title#1{\begin{center}{\Large\bf #1}\end{center}}
\def\author#1{\vskip 5mm \begin{center}{#1}\end{center}}
\def\address#1{\begin{center}{\it #1}\end{center}}

\newcommand{\bfr}{\begin{flushright}}
\newcommand{\efr}{\end{flushright}}

\begin{document}

\vspace*{-2cm}
\bfr{}\efr\vspace{-9mm}
\bfr{}\efr
\vspace{1cm}

\title{Soliton Equations Extracted from the Noncommutative Zero-Curvature Equation}\author{Takao KOIKAWA\footnote{E-mail: koikawa@otsuma.ac.jp}}
\vspace{1cm}
\address{
  School of Social Information Studies,
         Otsuma Women's University,\\
         Tama 206-0035,Japan\\
}
\vspace{2.5cm}
\abstract{ }
We investigate the equation where the commutation relation in 2-dimensional zero-curvature equation composed of the algebra-valued potentials is replaced by the Moyal bracket and the algebra-valued potentials are replaced by the non-algebra-valued ones with two more new variables. We call the 4-dimensional equation the noncommutative zero-curvature equation. We show that various soliton equations are derived by the dimensional reduction of the equation.

\newpage
\setcounter{page}{2}
\section{Introduction}

The Moyal algebra seems to have a rich structure in it\cite{Moy}.  It was first considered as an alternate way of the quantization. The parameter appearing there was the Planck $h$. The vanishing limit of the parameter in the Moyal bracket, which means taking the classical limit of the quantum theory, leads to the Poisson bracket. In this paper we shall study the formulation of the soliton equations derived from the equation including the Moyal algebra, which would help us to understand the implication of the algebra. We shall introduce a parameter through the Moyal algebra into the present soliton formulation, but it is not assumed to be the Planch $h$.  We shall study the role of the parameter in the present formulation and show that the role of the parameter is simply the lattice spacing of the discrete soliton equation case and the coefficients of higher derivative terms in the continuous soliton equation case. The comparison of the role of the parameter in the Moyal quantizaiton with that in the soliton formulation is quite interesting. In the Moyal quantization, the energy levels are discrete by the order of the Planck $h$, while in the discrete soliton equation, or the completely integrable case, the space is discrete by the order of the parameter.

It has been well known that the soliton equations are formulated as the two dimensional zero-curvature equation and the divergenceless equation.
\begin{eqnarray}
\frac{\partial A_{\nu}}{\partial x_{\mu}}-
\frac{\partial A_{\mu}}{\partial x_{\nu}}+[A_{\mu},A_{\nu}]&=&0,\\
\frac{\partial A_{\mu}}{\partial x_{\mu}}&=&0.
\end{eqnarray}
where $A_{\mu}=A_{\mu}(x_0,x_1)$($\mu=0,1$) are $sl(N,C)$
valued potentials. When the potentials include the spectral parameter, this is the inverse scattering method\cite{AKNS,AKS}. The potentials are algebra-valued functions of two variables $x_0$ and $x_1$. The Lax pair formulation can also be viewed to fall into this category when the potentials include the differential operators instead of the spectral parameter. In the previous papers\cite{Koi1,Koi2} we considered the ''zero-curvature equation" where the ordinary commutation relation appearing in the equation is replaced by the Moyal bracket and the potentials are the functions of 4 variables, $x_0$, $x_1$ and new variables $x$ and $p$, which makes the equation the 4-dimensional one. In order to distinguish the equation from the 2-dimensional zero-curvature equation, we call the 4-dimensional equation the noncommutative zero-curvature equation. The loss of the matrix form of the potentials accounts for the increase of the variables. In the previous papers, we considered the dimensional reduction of the equation and showed how the soliton equations on the one-dimensional lattce and their continuous limits are extracted from the noncommutative zero-curvature equation. Included were the Toda lattice equation, the Bogomolny equation\cite{Bogo}, and KM equation\cite{KM}. In these equations, the parameter sneaked in through Moyal bracket plays the role of the one-dimensional lattice spacing. Their continuous counterparts were also obtained by simply taking the vanishing limit of the parameter appearing in the definition of the Moyal bracket. We thus showed that the Moyal bracket formulation of the soliton equations comprise both the discrete and continuous soliton equations with a parameter interpolating them. The idea obtaining those equations exemplified above were based on the expansion of the functions of $x$ and $p$ in terms of ${\rm e}^p$ and this implies to extract the $su(N)$ structure from the Moyal bracket. In the previous papers we gave the coefficient functions a priori and did not show how the potentials leading to the various soliton equations are determined. We did not show how the KdV equation, which is one of the most well known soliton equations, can be derived in the previous papers
either.

In the present paper,  we shall discuss the reduction of the noncommutative zero-curvature equation to derive the 2-dimensional soliton equations. We shall show how the coefficients when the potentials are expanded in terms of ${\rm e}^p$ are to be determined. They are the reduction leading to the equations relevant to the Toda lattice equation. In order to point out the extensive validity of the extraction of the soliton equations from the noncommutative zero-curvature equation, we shall investigate the dimensional reduction by the power series expansion by the variable $p$, which also leads to the typical soliton equation known as the KdV equation and its higher order equations. 

This paper is constructed as follows. Next section is devoted to the summary of the definitions, notations and formula which we shall use in the following sections. We shall discuss the KdV and MKdV hierarchy equations in section 3. In section 4 we show the expansion of potentials in terms of ${\rm e}^p$ leads to the Toda lattice equation, Bogomolny equation and KM equation. The difficulty in obtaining the hierarchy of Toda lattice equation is also discussed in this section. The last section would be devoted to the summary and the discussion.

\section{Formulation}

We first recapitulate the notation of the Moyal algebra. The star product for functions $f=f(x,p)$ and $g=g(x,p)$  is defined by
\begin{equation}
f\star g = \exp \Bigg[ \kappa
\Bigg(\frac{\partial~}{\partial x}\frac{\partial~}{\partial\tilde p}-\frac{
\partial~}{\partial p}\frac{\partial~}{\partial\tilde x}\Bigg)
\Bigg] f({\bf x}) g({\bf \tilde x}) \vert_{{\bf x} = {\bf\tilde x}},
\label{eq:star1}
\end{equation}
where ${{\bf x}=(x,p)}$ and ${{\bf \tilde x}=(\tilde x,\tilde p)}$ and they are set equal after the derivatives are taken\cite{Str}. Here $\kappa$ is the parameter of which the physical meaning is not specified yet but would be discussed later.
The Moyal bracket is defined by using the star product as
\begin{equation}
\{f,g\}=\frac{f \star g - g \star f}{2 \kappa},
\end{equation}
which turns out to be the Poisson bracket of $f$ and $g$ in the vanishing limit of $\kappa$.

In the following sections, we expand the functions of $x$ and $p$ in terms of the powers of $p$ in the next section and ${\rm e}^p$ in section 4. When the functions are expanded in terms of ${\rm e}^p$, the Moyal bracket reads
\begin{eqnarray}
&{}& \{e^{mp}f,e^{np}g\} \nonumber \\
&=&\frac{1}{2\kappa}\exp\{(m+n)p\}\Bigg[f(x+n\kappa)g(x-m\kappa) - g(x+m\kappa)f(x-n\kappa)\Bigg],
\label{eq:bra}
\end{eqnarray}
where $m$ and $n$ are integers.
We can naturally obtain the $\kappa \to 0$ limit of the bracket:
\begin{equation}
\lim_{\kappa \to 0} \{e^{mp}f,e^{np}g\} = \Bigg(n\frac{df}{dx}g-m\frac{dg}{dx}f\Bigg)\exp\{(m+n)p\}.
\label{eq:lim}
\end{equation}

The soliton equations are known to be formulated as the 2-dimensional zero-curvature equation which emerges as a compatibility condition of two linear equations. The well known inverse scattering method and the Lax equation fall into the formulation. On the other hand, it was argued that the Moyal bracket can reproduce $su(N)$ algebra when we specify the geometry of the base space where the potentials are defined\cite{Hop,FFZ,FZ}. This fact motivated us to propose the equation which replaces the commutation relation in the zero-curvature equation with the Moyal bracket. The potentials are no more algebra valued. Instead they are now functions of new variables $x$ and $p$ besides old variables $x_0$ and $x_1$, and so they are the 4-dimensional fields. The equation reads
\begin{equation}
\frac{\partial A_{\nu}}{
\partial x_{\mu}}-\frac{\partial A_{\mu}}{\partial x_{\nu}}+\{A_{\mu},A_{\nu}\}=0,
\label{eq:0ceq}
\end{equation}
where $A_{\mu}=A_{\mu}(x_0, x_1;x, p),(\mu=0,1)$. Here we assume the potentials do not include either the differential operators nor the spectral parameter, but include a parameter coming in through the Moyal braket. We shall call this 4-dimensinal equation the noncommutative zero-curvature equation to distinguish this from the 2-dimensional zero-curvature equation. The noncommutative zero-curvature equation seems to have a rich algebraic structure despite that the potentials are not the algebra-valued. It is well known that many of the 2-dimensional soliton equations are formulated by the inverse scattering problem, and they are characterized by the algebra.
What we shall show in this paper is essentially the reduction of the noncommutative zero-curvature equation to derive the 2-dimensional soliton equations. This would turn out to reveal the rich algebraic and geometric structure of the noncommutative zero-curvature equation. 

In order to show that the well known 2-dimensional soliton equations are derived from the noncommutative zero-curvature equation, we first assume the $x_0$-independence or $x_1$-independence and then expand the potentials of the remaining 3 variables in terms of power series of the variable $p$ or ${\rm e}^p$ and determine coefficient functions by substituting them into the noncommutative zero-curvature equation. We thus obtain the 2-dimensional equations.

\section{KdV hierarchy equations}

In this section we make the power expansion of the potentials in terms of the variable $p$, and show that the specification of the expansions of potentials and their substitution into the noncommutative zero-curvature equation determines the coefficient functions of the expansion and so it leads to the specific soliton equation. We also assume that the potentials are independent of either $x_1$ or $x_0$. We first consider the $x_1$-independent case. We shall denote the potentials by $f$ and $g$ to avoid the troublesome suffix of the potentials:
\begin{eqnarray}
f&=&A_1(t;x,p),
\label{eq:def_f}
\\
g&=&A_0(t;x,p),
\label{eq:def_g}
\end{eqnarray}
where $t=x_0$. 
Then the Eq.(\ref{eq:0ceq}) reads
\begin{equation}
f_t=\{f, g \},
\label{eq:sol_eq}
\end{equation}
where $f_t={\partial f}/{\partial t}$. When we assume that the functions are $x_0$-independent and denote $x_1$ as $s$: 
\begin{eqnarray}
f&=&A_1(s;x,p),
\label{eq:def_f2}
\\
g&=&A_0(s;x,p),
\label{eq:def_g2}
\end{eqnarray}
we obtain
\begin{equation}
g_s=-\{f, g \}.
\label{eq:sol_eq2}
\end{equation}
We can extract the soliton equations from the Eq. (\ref{eq:sol_eq}) or (\ref{eq:sol_eq2}) by the power series expansions of $f$ and $g$ as:
\begin{eqnarray}
f(t;x,p)&=&\sum_{n=0}^N p^n f_n(t;x),
\label{eq:pexp_f}
\\
g(t;x,p)&=&\sum_{m=0}^M p^n g_n(t:x),
\label{eq:pexp_g}
\end{eqnarray}
where we assume that $f_N=1$ and $g_M=1$. The specification of $M$ and $N$ in the Eqs.(\ref{eq:pexp_f}) and (\ref{eq:pexp_g}) determines the specific equation.

As the first example, we shall see how the KdV equation and the Boussinesque equation are obtained. They are derived in a pair. Expanding $f=f(t;x,p)$ and $g=g(t;x,p)$ as
\begin{eqnarray}
f(t;x,p)&=&p^2+u(t;x),
\\
g(t;x,p)&=&p^3+p^2 g_2(t;x)+ p g_1(t;x)+g_0(t;x),
\end{eqnarray}
we insert them into Eq.(\ref{eq:sol_eq}). Then the coefficient functions $g_k$(k=0,1,2) are determined in terms of $u$ and its $x$-derivatives by comparing the both sides of the equation at each order of power of $p$.  Eq.(\ref{eq:sol_eq}) thus leads to the KdV equation:
\begin{equation}
u_t=\frac{3}{2}u'u+\kappa^2u'''.
\end{equation}
where $u'={\partial u}/{\partial x}$. 

When we assume the $t$ independence of the potentials and expand them as 
\begin{eqnarray}
f(s;x,p)&=&p^2+u(s;x),
\\
g(s;x,p)&=&p^3+p^2 g_2(s;x)+ p g_1(s;x)+g_0(;x),
\
\end{eqnarray}
we obtain the Boussinesque equation by inserting them into Eq.(\ref{eq:sol_eq2}):
\begin{equation}
u_{ss}+(uu'+\kappa^2u''')'=0.
\label{eq:Bouss}
\end{equation}
We should note that these soliton equations are obtained by the same specification of $M$ and $N$. The only difference is the assumption on which variables the potentials depend. In this sense they are obtained in a pair.

The above observation motivates us to study more extensive and systematic study of the noncommutative zero-curvature equation to derive the soliton equations. We shall show that the following power series expansions of the potentials lead to the KdV hierarchy equations:
\begin{eqnarray}
f(t;x,p)&=&p^2+u(t;x),
\label{eq:KdVExp1}
\\
g(t;x,p)&=&\sum_{k=0}^{2N+1} p^kg_k(t;x).
\label{eq:KdVExp2}
\end{eqnarray}
Here and hereafter we denote $g_k(t;x)$ as $g_k$. Substituting above expansions into the Moyal bracket, we are left with
\begin{eqnarray}
\{f,g\}&=&\sum_{m=0}^{N} \kappa^{2m} \sum_{k=0}^{2(N-m)}{k+2m+1 \choose 2m+1} p^k u^{(2m+1)} g_{2k+2m+1} 
\nonumber \\
& &{}-2 \sum_{k=1}^{2N+1} p^k \frac{\partial g_{n-1}}{\partial x} \nonumber \\
&=&\sum_{m=0}^{N} \kappa^{2m} u^{(2m+1)} g_{2m+1} \nonumber \\
& &{}+\sum_{k=1}^{2N+1} p^k \bigl\{ \sum_{m=0}^{[\frac{2N-k}{2}]} \kappa^{2m} {k+2m+1 \choose 2m+1} u^{(2m+1)}g_{k+2m+1}-2\frac{\partial g_k}{\partial x} \bigr\} \nonumber \\
& &{}-2 p^{2N+1} \frac{\partial g_{2N}}{\partial x},
\end{eqnarray}
where $[a]$ is the integer number which does not exceed $a$. 
The comparison of the both sides of Eq.(\ref{eq:sol_eq}) at each order of power of $p$ shows that
\begin{eqnarray}
u_t&=&\sum_{m=1}^N \kappa^{2m}u^{(2m+1)}g_{2m+1}, \label{eq:KdVH}\\
\frac{\partial g_{k-1}}{\partial x}&=&\frac{1}{2}\sum_{m=0}^{[\frac{2N-k}{2}]} \kappa^{(2m)}{k+2m+1 \choose 2m+1}u^{(2m+1)} g_{k+2m+1},\\
& &\qquad \qquad \qquad \qquad \qquad (k=1,2,\cdots ,2N) \nonumber \\
\frac{\partial g_{2N}}{\partial x}&=&0.
\end{eqnarray}
where $u^{(k)}={\partial^k u}/{\partial x_k}$. The first equation is the soliton equation of the KdV hierarchy series. The functions $g_l$ appearing on the RHS are completely determined by other equations. Those equations solve $g_{2k}$ as
\begin{equation}
g_{2k}=0,(k=1,2,\cdots ,N).
\end{equation}
On the other hand, $g_l$ with odd number of suffix $l$ are determined by
\begin{equation}
\frac{\partial g_{2k-1}}{\partial x}=\frac{1}{2}\sum_{m=0}^{N-k} \kappa^{(2m)}{2k+2m+1 \choose 2m+1}u^{(2m+1)} g_{2(k+m)+1},\\
\end{equation}
which shows that $g_{2k-1}$ is expressed in terms of ${g_{2l-1}}(l>k)$ and the derivatives of $u(t;x)$, and so we can obtain all the $g_{2k-1}$ recursively when the equations are combined with the condition $g_{2N+1}=1$. By substituting the results, we obtain the KdV hierarchy equations. Though we have not yet proven that all the series are completely the same as the KdV known hierarchy series at each order, we have obtained completely identical equations for the first 4 equations. Here we list them below. Denoting the RHS of Eq.(\ref{eq:KdVH}) by ${\cal K}_{2N+1}$ for each $N$, we get
\begin{eqnarray}
u_t&=&u^{(1)}={\cal K}_1,\\
u_t&=&\frac{3}{2}u u^{(1)}+\kappa^2u^{(3)}={\cal K}_3,\label{eq:KdV}\\
u_t&=&\frac{15}{8}u^2u^{(1)}+5 \kappa^2 u^{(1)} u^{(2)}+\frac{5}{2} \kappa^2 u u^{(3)}+\kappa^4 u^{(5)}={\cal K}_5,\\
u_t&=&\frac{35}{16}u^3u^{(1)}+\frac{35}{8}\kappa^2(u^{(1)})^3+\frac{35}{2}\kappa^2uu^{(1)}u^{(2)}+\frac{21}{2}\kappa^4u^{(1)}u^{(4)} \nonumber \\
&&{}+\frac{35}{8}\kappa^2 u^2 u^{(3)}+\frac{35}{2}\kappa^4 u^{(2)} u^{(3)}+\frac{7}{2}\kappa^4 uu^{(5)}+\kappa^6u^{(7)}={\cal K}_7.\\
\end{eqnarray}
Here we note that the power of $\kappa$ measures the order of derivatives. Let us introduce an integro-differential operator defined by
\begin{equation}
{\cal O}=\kappa^2 D^2+u+u^{(1)}I,
\end{equation}
where $D=\partial/\partial x$ and $I=\int^x dx$. The operator relates ${\cal K}_{2k+1}$ and ${\cal K}_{2k-1}$. Then, by using the operator, we can rewrite the above series as
\begin{eqnarray}
u_t&=&{\cal K}_1,\\
u_t&=&{\cal K}_3={\cal O}{\cal K}_1,\\
u_t&=&{\cal K}_5={\cal O}{\cal K}_3={\cal O}^2{\cal K}_1,\\
u_t&=&{\cal K}_7={\cal O}{\cal K}_5={\cal O}^3{\cal K}_1.
\end{eqnarray}
When $\kappa=1/2$, the integro-differential operator is identical to that found in the Hamiltonian formulation\cite{AKS} for the KdV equation given by
\begin{equation}
u_t=6uu^{(1)}+u^{(3)}.
\end{equation}
This is obtained by setting the same value $1/2$ to $\kappa$ in Eq.(\ref{eq:KdV}) together with the scaling of $t$. We have thus extracted the KdV hierarchy equations from the Moyal bracket zero-curvature equation by assuming the expansions by (\ref{eq:KdVExp1}) and (\ref{eq:KdVExp2}).

We next study different type of the expansions expecting to get the MKdV hierarchy equations. Instead of the expansions given by (\ref{eq:KdVExp1}) and (\ref{eq:KdVExp2}) we exploit other series obtained by the following expansions:
\begin{eqnarray}
f(t;x,p)&=&p+u(t;x),
\label{eq:MKdVExp1}
\\
g(t;x,p)&=&\sum_{k=1}^{M} p^kg_k(t;x),
\label{eq:MKdVExp2}
\end{eqnarray}
with the condition that $g_M=1$.
In the similar way as the KdV equation, we obtain the MKdV equation for M=3 case. We shall study the higher series of equations in the similar way as in the KdV hierarchy equations.  Two of the series are obtained as
\begin{eqnarray}
u_t&=&\kappa^2u^{(3)}+3u^2u'=\tilde{\cal K}_3,
\\
u_t&=&\kappa^5 u^{(5)}+20\kappa^2uu'u^{(2)}+10\kappa^2 u^2 u^{(3)}+5u^4u'=\tilde{\cal K}_5.
\end{eqnarray}
In order to see whether these are the first few MKdV hierarchy series, they are to be compared with the known MKdV series by adjusting the value of $\kappa$. The result shows that the terms appearing in the above results are the same but the coefficients can not be adjusted to be the same as the known MKdV series. They are slightly different, which should not be overlooked. In contrast with the KdV case, we can not construct the operator connecting $\tilde{\cal K}_n$. According to the Sato theory, the ${\cal K}_n$ in the KdV  hierarchy equations are regarded as the generators of the infinite number of symmetries of the KdV equations. Then we can write the equations by introducing the infinite number of times $t_{2n+1}$ by
\begin{equation}
u_{t_{2n+1}}={\cal K}_{2n+1},(n=1,2,\cdots).
\end{equation}  
They should be compatible, which requires
\begin{equation}
\frac{\partial {\cal K}_{2n+1}}{\partial t_{2m+1}}=\frac{\partial {\cal K}_{2m+1}}{\partial t_{2n+1}}.
\end{equation}  
These are satisfied by the KdV series obtained above,  but the similar equations do not hold for $\tilde{\cal K}$ of the MKdV hierarchy equations. From these reasons we conclude the expansions (\ref{eq:MKdVExp1}) and (\ref{eq:MKdVExp2}) do not give rise to the MKdV hierarchy equations, though the first non-trivial equation is the MKdV equation.

\section{Toda Lattice hierarchy equations}

In the previous section, the power series expansion of the potentials are found to lead to the KdV hierarchy equations. Here we exploit other expansions which lead to the Toda lattice equation and other relevant equations. One may expect to obtain the Toda lattice hierarchy equations like the KdV hierarchy equations which are supposed to incorporate the interactions of a particle on one lattice site with other particles sitting on other sites farther than the neighboring sites. In this section we shall study the case where the potentials are expressed as
\begin{eqnarray}
A_0&=&\sum_k {\rm e}^{kp}a_k,
\\
A_1&=&\sum_k {\rm e}^{kp}b_k.
\end{eqnarray}
Another assumption is that $A_i=A_i(x_0=t;x,p)$,(i=1,2). Then the Eq.(\ref{eq:0ceq}) reads
\begin{equation}
\frac{{\rm d}A_1}{{\rm d}t}=\{A_1,A_0\}.
\label{eq:0ceqt}
\end{equation}
This might be replaced by the following equation for the potentials given by $A_i=A_i(x_1=r;x,p)$,(i=1,2).  Eq.(\ref{eq:0ceq}) reads
\begin{equation}
\frac{{\rm d}A_0}{{\rm d}r}=-\{A_1,A_0\}.
\label{eq:0ceqr}
\end{equation}

We shall first expand the potentials as:
\begin{eqnarray}
A_0&=&a_0(t;x)+a_{-1}(t;x){\rm e}^{-p}+a_1(t;x){\rm e}^p,
\label{eq:TodaExp1}
\\
A_1&=&b_0(t;x)+b_{-1}(t;x){\rm e}^{-p}+b_1(t;x){\rm e}^p,
\label{eq:TodaExp2}
\end{eqnarray}
and substitute them into the Eq.(\ref{eq:0ceqt}) to obtain at each power of ${\rm e}^p$
\begin{eqnarray}
0&=&a_{-1}(x-\kappa)b_{-1}(x+\kappa)-b_{-1}(x-\kappa)a_{-1}(x+\kappa),
\label{eq:Toda1}
\\
\frac{{\rm d} b_{-1}(x)}{{\rm d}t} &=& -\frac{1}{2\kappa}\big[a_{-1}(x)(b_{0}(x+\kappa)-b_{0}(x-\kappa))\nonumber \\
& & {}-b_{-1}(x)(a_{0}(x+\kappa)- a_{0}(x-\kappa)) \big],
\label{eq:Toda2}
\\
\frac{{\rm d} b_{0}(x)}{{\rm d}t} &=& -\frac{1}{2\kappa}\big[(a_{-1}(x+\kappa)b_{1}(x+\kappa)-b_{1}(x-\kappa)a_{-1}(x-\kappa))\nonumber \\
& & {}-(a_{1}(x+\kappa)b_{-1}(x+\kappa)-b_{-1}(x-\kappa)a_{1}(x-\kappa)) \big],
\label{eq:Toda3}
\\
\frac{{\rm d} b_{1}(x)}{{\rm d}t} &=& \frac{1}{2\kappa}\big[a_{1}(x)(b_{0}(x+\kappa)-b_{0}(x-\kappa))\nonumber \\
& & {}-b_{1}(x)(a_{0}(x+\kappa)- a_{0}(x-\kappa)) \big],
\label{eq:Toda4}
\\
0&=&a_{1}(x-\kappa)b_{1}(x+\kappa)-b_{1}(x-\kappa)a_{1}(x+\kappa).
\label{eq:Toda5}
\end{eqnarray}
Here and hereafter we abbreviate $t$-dependence of the functions and denote them like $a_i$ and $b_i$ unless the specification is necessary. In order to solve these equations we assume that $b_{\pm 1}$ are proportional to $a_{\pm 1}$:
\begin{equation}
b_{\pm 1}=\gamma_{\pm 1}a_{\pm 1},
\label{eq:prop}
\end{equation}
where $\gamma_{\pm 1}$ are non-zero constants. These solve Eqs.(\ref{eq:Toda1}) and (\ref{eq:Toda5}) and the remaining equations are written as
\begin{eqnarray}
\rm{ln} a_{\pm 1}(x)&=& \mp \frac{1}{2\kappa}[(a_0(x+\kappa)-a_0(x-\kappa))\nonumber  \\
& & {}-\frac{1}{\gamma_{\pm 1}}(b_0(x+\kappa)-b_0(x-\kappa))],
\\
\frac{{\rm d} b_{0}(x)}{{\rm d}t} &=& -\frac{1}{2\kappa}(\gamma_1-\gamma_{-1})[a_{-1}(x+\kappa)a_{1}(x+\kappa) \nonumber \\
& & {}- a_{1}(x-\kappa)a_{-1}(x-\kappa)].
\end{eqnarray}
Introducing $\rho(t;x)$ by 
\begin{equation}
\rho(t;x)=-{\rm ln}a_{-1}(t;x)a_{1}(t;x),
\end{equation}
we obtain from above equations
\begin{equation}
\frac{{\rm d}^2 \rho(t;x)}{{\rm d}t^2}=-\frac{(\gamma_1-\gamma_{-1})^2}{\gamma_1 \gamma_{-1}}\frac{1}{(2\kappa)^2}\big( {\rm e}^{-\rho(x-2\kappa)}-2{\rm e}^{-\rho(x)}+{\rm e}^{-\rho(x+2\kappa)}\big).
\end{equation}
When we denote the discrete space by $x=\kappa n$,(n=1,2,$\cdots$) and write $\rho(t;x)$ as $\rho_n(t)$, we get the Toda lattice equation of which the spacing is given by $\kappa$:
\begin{equation}
\frac{{\rm d}^2 \rho_n(t)}{{\rm d}t^2}=-\frac{(\gamma_1-\gamma_{-1})^2}{\gamma_1 \gamma_{-1}}\frac{1}{(2\kappa)^2}\big({\rm e}^{-\rho_{n-2}(t)}-2{\rm e}^{-\rho_n(t)}+{\rm e}^{-\rho_{n+2}(t)}\big).
\label{eq:TodaEq}
\end{equation}

In the similar way, when we assume that the potentials are the functions of not $t$, $x$ and $p$ but $r$, $x$ and $p$ in Eqs.(\ref{eq:TodaExp1}) and (\ref{eq:TodaExp2}), we obtain from Eq.(\ref{eq:0ceqr})
\begin{equation}
\frac{{\rm d}^2 \rho_n(r)}{{\rm d}r^2}=-(\gamma_1-\gamma_{-1})^2\frac{1}{(2\kappa)^2}\big( {\rm e}^{-\rho_{n-2}(r)}-2{\rm e}^{-\rho_n(r)}+{\rm e}^{-\rho_{n+2}(r)}\big),
\label{eq:BogoEq}
\end{equation}
which may be called the Bogomolny equation when the overall sign of the RHS is set to be positive, which might be realized by choosing the pure imaginary $\gamma_i$. Eqs.(\ref{eq:TodaEq}) and (\ref{eq:BogoEq}) are obtained in pairs as in the KdV and Boussinesque equations in the previous section. But in the present case there is no significant difference between them, because the potentials are expanded in a symmetric way in deriving these equations. Actually Eq.(\ref{eq:TodaEq}) is also regarded as the Bogomolny equation by regarding the time variable as the radial variable and choosing $\gamma_0$ and $\gamma_1$ with different signs. We can take the limit that $\kappa \to 0$ in the above equations. Then we get the continuous Toda lattice equation and the continuous Bogomolny equations, respectively:
\begin{eqnarray}
\frac{\partial^2 \rho}{\partial t^2}&=&\frac{\partial^2 {\rm e}^{-\rho}}{\partial x^2},
\\
\frac{\partial^2 \rho}{\partial r^2}&=&-\frac{\partial^2 {\rm e}^{-\rho}}{\partial x^2}.
\end{eqnarray}
They were shown to have an infinite number of conserved currents\cite{Koi0,Koi2}. This means that the noncommutative zero-curvature equation leads to the completely integrable system under the expansion of the potentials in terms of ${\rm e}^p$ irrespective of the value of $\kappa$.

We next expand the potentials in an asymmetric way as
\begin{eqnarray}
A_0&=&a_0(t;x)+a_{-1}(t;x){\rm e}^{-p}+a_1(t;x){\rm e}^p
\nonumber \\
&&{}+a_{-2}(t;x){\rm e}^{-2p}+a_2(t;x){\rm e}^{2p},
\label{eq:TodaHExp1}
\\
A_1&=&b_0(t;x)+b_{-1}(t;x){\rm e}^{-p}+b_1(t;x){\rm e}^p,
\label{eq:TodaHExp2}
\end{eqnarray}
where we assume that $a_{\pm 2}$ are not zero. By substituting these into Eq.(\ref{eq:0ceqt}) we obtain at each order of power of ${\rm e}^p$;
\begin{eqnarray}
0&=&a_{\mp 2}(x\mp \kappa)b_{\mp 1}(x \pm 2\kappa)-b_{\mp 1}(x \mp 2\kappa)a_{\mp 2}(x \pm \kappa),
\label{eq:KM1}
\\
0&=&a_{\mp 1}(x \mp \kappa)b_{\mp 1}(x \pm \kappa)-b_{\mp 1}(x \mp \kappa)a_{\mp 1}(x \pm \kappa)
\nonumber \\
&&{}+a_{\mp 2}(x)(b_0(x \pm2 \kappa)-b_0(x \mp 2\kappa)),
\label{eq:KM2}
\\
\frac{{\rm d}b_{\mp 1}(x)}{{\rm d}t} &=& \mp \frac{1}{2\kappa}\big[a_{\mp 1}(x)(b_{0}(x+\kappa)-b_{0}(x-\kappa))
\nonumber \\
&&{}-b_{\mp 1}(x)(a_{0}(x+\kappa)- a_{0}(x-\kappa))
\nonumber \\
&&{}+(a_{\mp 2}(x \pm \kappa)b_{\pm 1}(x \pm 2\kappa)-b_{\pm 1}(x \mp 2\kappa)a_{\mp 2}(x \mp \kappa)) \big],
\label{eq:KM3}
\\
\frac{{\rm d} b_{0}(x)}{{\rm d}t} &=& -\frac{1}{2\kappa}\big[a_{-1}(x+\kappa)b_{1}(x+\kappa)-b_{1}(x-\kappa)a_{-1}(x-\kappa))\nonumber \\
&-& (a_{1}(x+\kappa)b_{-1}(x+\kappa)-b_{-1}(x-\kappa)a_{1}(x-\kappa)) \big].
\label{eq:KM4}
\end{eqnarray}
The last equation is the same as Eq.(\ref{eq:Toda3}) in the previous case, while other equations are subject to the changes due to the inclusion of $a_{\pm 2}(x)$ terms to $A_0$. We assume as before that
\begin{equation}
b_{\pm 1}(x)=\gamma_{\pm 1}a_{\pm 1}(x),
\end{equation}
where $\gamma_{\pm 1}$ are constants. Then the Eqs.(\ref{eq:KM1}) and (\ref{eq:KM2}) lead to
\begin{eqnarray}
a_{\pm2}(x+\kappa)a_{\pm 1}(x-2\kappa)-a_{\pm 1}(x + 2\kappa)a_{\pm 2}(x - \kappa)&=&0,
\\
a_{\pm 2}(x)(b_0(x+2 \kappa)-b_0(x - 2\kappa))&=&0.
\end{eqnarray}
These are solved by
\begin{eqnarray}
a_{\pm2}(x)&=&k_{\pm2}a_{\pm 1}(x-\kappa)a_{\pm 1}(x + \kappa),
\\
b_0(x)&=&C,
\end{eqnarray}
where $k_{\pm2}$ and $C$ are constants. By substituting these into Eqs.(\ref{eq:KM3}) and (\ref{eq:KM4}), we can rewrite them as 
\begin{eqnarray}
\frac{{\rm d}{\rm ln} a_{\mp 1}(x)}{{\rm d}t} &=& \pm \frac{1}{2\kappa}\big[ (a_{0}(x+\kappa)- a_{0}(x-\kappa))+\frac{k_{\pm 2} \gamma_{\mp 1}}{\gamma_{\pm 1}}\times
\nonumber \\
&&{}(a_{-1}(x + 2\kappa)a_{1}(x +2\kappa)-a_{-1}(x-2\kappa)a_{1}(x-2\kappa)) \big],
\\
0 &=& -\frac{(\gamma_{-1}-\gamma_1)}{2\kappa} \times
\nonumber \\
&&{}(a_{-1}(x+\kappa)a_{1}(x+\kappa)-a_{1}(x-\kappa)a_{-1}(x-\kappa)).
\end{eqnarray}
In order to obtain the non-trivial solutions, we require that $\gamma_1=\gamma_{-1}$. Introducing $\rho(t;x)$ by 
\begin{equation}
\rho(t;x)=-{\rm ln}a_{-1}(t;x)a_{1}(t;x),
\end{equation}
as in the preceding example, we obtain 
\begin{equation}
\frac{{\rm d} \rho(x)}{{\rm d}t}=-\frac{(k_2 -k_{-2})}{2 \kappa}({\rm e}^{-\rho(x+2\kappa)}-{\rm e}^{-\rho(x-2\kappa)}),
\end{equation}
which is the well-known KM equation.

When we assume the $r$ dependence instead of the $t$ dependence in Eqs.(\ref{eq:TodaExp1}) and (\ref{eq:TodaExp2}) and use Eq.(\ref{eq:0ceqr}) instead of Eq.(\ref{eq:0ceqt}), we can not derive any non-trivial equation under the assumptions $b_{\pm 1}(r;x)=\gamma_{\pm 1}a_{\pm 1}(r;x)$. Therefore there is no pairwise equation with the KM equation.

The next example of the expansions of the potentials are given by
\begin{eqnarray}
A_0&=&a_0(t;x)+a_{-1}(t;x){\rm e}^{-p}+a_1(t;x){\rm e}^p
\nonumber \\
&&{}+a_{-2}(t;x){\rm e}^{-2p}+a_2(t;x){\rm e}^{2p}
\nonumber \\
&&{}+a_{-3}(t;x){\rm e}^{-3p}+a_3(t;x){\rm e}^{3p},
\label{eq:TodaExp1c}
\\
A_1&=&b_0(t;x)+b_{-1}(t;x){\rm e}^{-p}+b_1(t;x){\rm e}^p,
\label{eq:TodaExp2c}
\end{eqnarray}
The substitution of the expansions into Eq.(\ref{eq:0ceqt}) yields the equations determining $a_k(t;x)$ and $b_k(t;x)$ at each order of power of ${\rm e}^p$ as in the previous examples. The newly appearing and the modified equations are given by
\begin{eqnarray}
0&=&a_{\mp 3}(x\mp \kappa)b_{\mp 1}(x \pm 3\kappa)-b_{\mp 1}(x \mp 3\kappa)a_{\mp 3}(x \pm \kappa),
\label{eq:KM1a}
\\
0&=&a_{\mp 2}(x\mp \kappa)b_{\mp 1}(x \pm 2\kappa)-b_{\mp 1}(x \mp 2\kappa)a_{\mp 2}(x \pm \kappa)
\nonumber \\
&&{}-a_{\mp 3}(x)(b_0(x \pm 3\kappa)-b_0(x \mp 3\kappa)),
\label{eq:KM2a}
\\
0&=&a_{\mp 1}(x \mp \kappa)b_{\mp 1}(x \pm \kappa)-b_{\mp 1}(x \mp \kappa)a_{\mp 1}(x \pm \kappa)
\nonumber \\
&&{}+a_{\mp 2}(x)(b_0(x \pm2 \kappa)-b_0(x \mp 2\kappa))
\nonumber \\
&&{}+a_{\mp 3}(x \pm \kappa)b_{\pm 1}(x \pm 3\kappa)-b_{\pm 1}(x \mp 3\kappa)a_{\mp 3}(x \mp \kappa).
\label{eq:KM3a}
\end{eqnarray}
The other equations at $O({\rm e}^{\pm p})$ and $O(1)$ are not subject to the modification even if we add the $a_{\pm 3}$ terms and so the Eqs.(\ref{eq:KM3}) and (\ref{eq:KM4}) still hold. We then assume as before that
\begin{equation}
b_{\pm 1}=\gamma_{\pm 1}a_{\pm 1}.
\end{equation}
Then the above three equations are reduced to 
\begin{eqnarray}
0&=&a_{\mp 3}(x\mp \kappa)a_{\mp 1}(x \pm 3\kappa)-a_{\mp 1}(x \mp 3\kappa)a_{\mp 3}(x \pm \kappa),
\\
0&=&\gamma_{\pm 1}(a_{\mp 2}(x\mp \kappa)a_{\mp 1}(x \pm 2\kappa)-a_{\mp 1}(x \mp 2\kappa)a_{\mp 3}(x \pm \kappa))
\nonumber \\
&&{}-a_{\mp 3}(x)(b_0(x \pm 3\kappa)-b_0(x \mp 3\kappa)),
\\
0&=&a_{\mp 2}(x)(b_0(x \pm2 \kappa)-b_0(x \mp 2\kappa))
\nonumber \\
&&{}+\gamma_{\pm 1}(a_{\mp 3}(x \pm \kappa)a_{\pm 1}(x \pm 3\kappa)-a_{\pm 1}(x \mp 3\kappa)a_{\mp 3}(x \mp \kappa)).
\end{eqnarray}
These are solved by
\begin{eqnarray}
a_{\pm 3}(x)&=&k_{\pm 3}a_{\pm 1}(x-2\kappa)a_{\pm 1}(x)a_{\mp 1}(x + 2\kappa),
\\
a_{\pm 2}(x)&=&k_{\pm 2}a_{\pm 1}(x-\kappa)a_{\mp 1}(x + \kappa),
\\
b_0(x)&=&C,
\end{eqnarray}
where $k_{\pm 3}$, $k_{\pm 2}$ and $C$ are constants. The fact that the Eqs.(\ref{eq:KM3}) and (\ref{eq:KM4}) are not subject to the modification from the previous KM equation case and the fact that $a_{\pm 2}(x)$ are the same solution as the KM equation case show that we obtain the same KM equation. Therefore in this case we can solve the equations without inconsistency, although there appears no new equation. 

\section{Summary and Discussions}

We have investigated the derivation the soliton equations from the noncommutative zero-curvature equation, which is based upon the various expansions of the potentials. It is intriguing that we can extract various soliton equations, which is known as the zero-curvature equations, from the noncommutative zero-curvature equation. We showed the KdV hierarchy equations in section 2 and Toda lattice equation and its related equations in section 3. Besides these equations we also studied other combinations of the expansions of the potentials. As for the MKdV equation, we obtained a series of equations. However we could not regard them as the MKdV series as we discussed in section 3. While the expansions adopted for the KdV hierarchy equations derive the zero-curvature equations from the noncommutative zero-curvature equation, the other choice of expansions could not extract MKdV series properly. This shows that arbitrary choice of the expansions of the potentials in the noncommutative zero-curvature equations does not necessarily lead to the soliton equation which is known to be the zero-curvature equation. For the moment, we have not found the definite criterion for choosing the proper expansions of the potentials leading to the soliton equations. 

As for the expansions in terms of ${\rm e}^p$, we also studied the equation obtained by the following expansions,  expecting to obtain the Toda lattice type equation which incorporates interaction of a particle on one-dimensional lattice with other particles on further sites than those of the Toda lattice equation:
\begin{eqnarray}
A_0&=&\sum_{k=-3}^3 {\rm e}^{kp}a_k(t;x),
\\
A_1&=&\sum_{k=-3}^3 {\rm e}^{kp}b_k(t;x).
\end{eqnarray}
As far as we assume that $b_i=\gamma_i a_i$,(i=1,2,3), we could not obtain any new equation. This is because the inclusion of new terms to $A_0$ and $A_1$ brings about more strict restrictions to the coefficient functions at each order of power of ${\rm e}^p$ in comparative with the Toda lattice equation case and so this does not allow for the emergence of the interaction besides those of the Toda lattice equation. This should be contrasted with the KdV case. The expansion in terms of ${\rm e}^p$ is regarded as the infinite series expansions in terms of $p$. The inclusion of other terms as above corresponds to the alternation of the coefficient values of the infinite series expansions. The fact that they are not allowed, though under our assumptions, might suggest the uniqueness of the Toda lattice equation, or the uniqueness of the completely integrable equation of this type.

\end{document}